\documentclass[11pt,a4paper]{article}
\usepackage[hyperref]{acl2019}
\usepackage{times}
\usepackage{latexsym}
\usepackage{graphicx}
\usepackage{url}
\usepackage{dblfloatfix}    

\aclfinalcopy 

\title{Visualizing Attention in Transformer-Based \\ Language Representation Models}

\author{Jesse Vig \\
  Palo Alto Research Center \\
    3333 Coyote Hill Road \\
    Palo Alto, CA 94304 \\
  {\tt jesse.vig@parc.com}}

\date{}

\begin{document}
\maketitle
\begin{abstract}
  We present an open-source tool for visualizing multi-head self-attention in Transformer-based language representation models. The tool extends earlier work by visualizing attention at three levels of granularity: the attention-head level, the model level, and the neuron level. We describe how each of these views can help to interpret the model, and we demonstrate the tool on the BERT model and the OpenAI GPT-2 model. We also present three use cases for analyzing GPT-2: detecting model bias, identifying recurring patterns, and linking neurons to model behavior.
\end{abstract}

\section{Introduction}

In 2018, the BERT (Bidirectional Encoder Representations from Transformers) language representation model achieved state-of-the-art performance across NLP tasks ranging from sentiment analysis to question answering \citep{Bert2018}. Recently, the OpenAI GPT-2 (Generative Pretrained Transformer-2) model  achieved state-of-the-art results across several language modeling benchmarks in a zero-shot setting \cite{gpt2}.

Underlying BERT and GPT-2 is the Transformer model, which uses a multi-head self-attention architecture \cite{transformer}. An advantage of using attention is that it can help interpret a model's decisions by showing how the model attends to different parts of the input \citep{align_translate, belinkov:2018:tacl}. Various tools have been developed to visualize attention in NLP models, ranging from attention-matrix heatmaps \citep{align_translate, neural_attention, rocktaschel2016reasoning} to bipartite graph representations \citep{visual_interrogation, Lee2017, seq2seqvisv1}.  A visualization tool designed specifically for the multi-head self-attention in the Transformer \citep{JonesViz} was introduced in \cite{transformer_arxiv} and released in the Tensor2Tensor repository \citep{tensor2tensor}. 

In this paper, we introduce a tool for visualizing attention in Transformer-based language representation models, building on the work of \cite{JonesViz}. We extend the existing tool in two ways: (1) we adapt it from the original encoder-decoder implementation to the decoder-only GPT-2 model and the encoder-only BERT model, and (2) we add two visualizations: the \textit{model view}, which visualizes all of the layers and attention heads in a single interface, and the \textit{neuron view}, which shows how individual neurons influence attention scores. 

\section{Visualization Tool}

We present an open-source\footnote{\url{https://github.com/jessevig/bertviz}} tool for visualizing multi-head self-attention in Transformer-based language representation models. The tool comprises three views: an attention-head view, a model view, and a neuron view. Below, we describe these views and demonstrate them on the OpenAI GPT-2 and BERT models. We also present three use cases for GPT-2 showing how the tool might provide insights on how to adjust or improve the model. A video demonstration of the tool is available at \url{https://youtu.be/187JyiA4pyk}. 


\subsection{Attention-head view}

\begin{figure*}[h]
    \includegraphics[width=\linewidth]{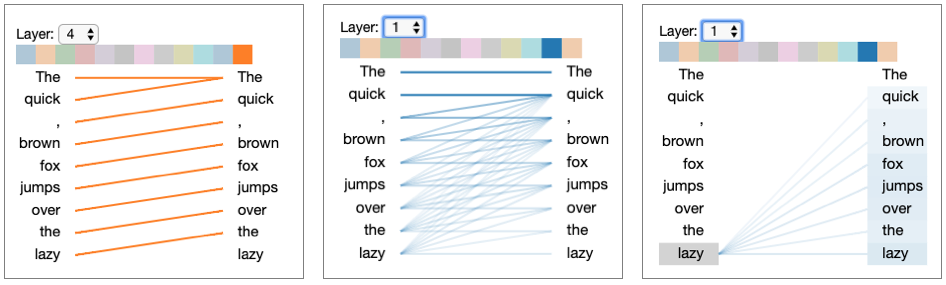}
    \caption{Attention head view for GPT-2, for the input text \textit{The quick, brown fox jumps over the lazy}. The left and center figures represent different layers / attention heads. The right figure depicts the same layer/head as the center figure, but with the token \textit{lazy} selected.}
    \label{fig:head_view_1_combined}

    \includegraphics[width=\linewidth]{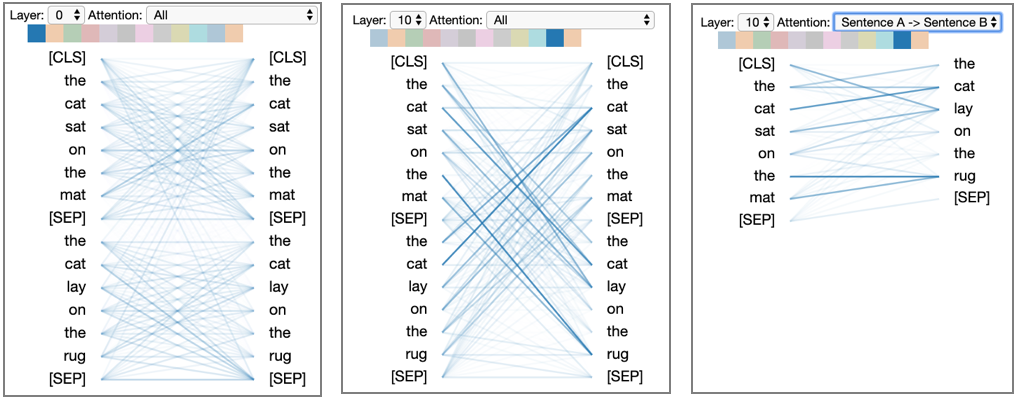}
    \caption{Attention head view for BERT, for inputs \textit{the cat sat on the mat} (Sentence A) and \textit{the cat lay on the rug} (Sentence B). The left and center figures represent different layers / attention heads. The right figure depicts the same layer/head as the center figure, but with \textit{Sentence A} $\rightarrow$ \textit{Sentence B} filter selected.}
    \label{fig:bert_heads}
    
\end{figure*}

\begin{figure*}
\includegraphics[width=\linewidth]{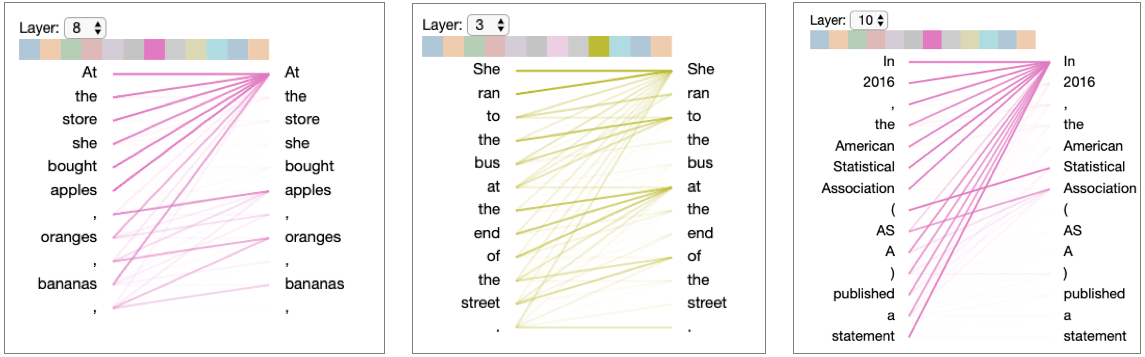}
    \caption{Examples of attention heads in GPT-2 that capture specific lexical patterns:  list items (left); prepositions (center); and acronyms (right). Similar patterns were observed in these attention heads for other inputs. (Attention directed toward first token is likely null attention, as discussed later.)}
    \label{fig:example_combined}
\end{figure*}

\begin{figure*}[h]
\includegraphics[width=\linewidth]{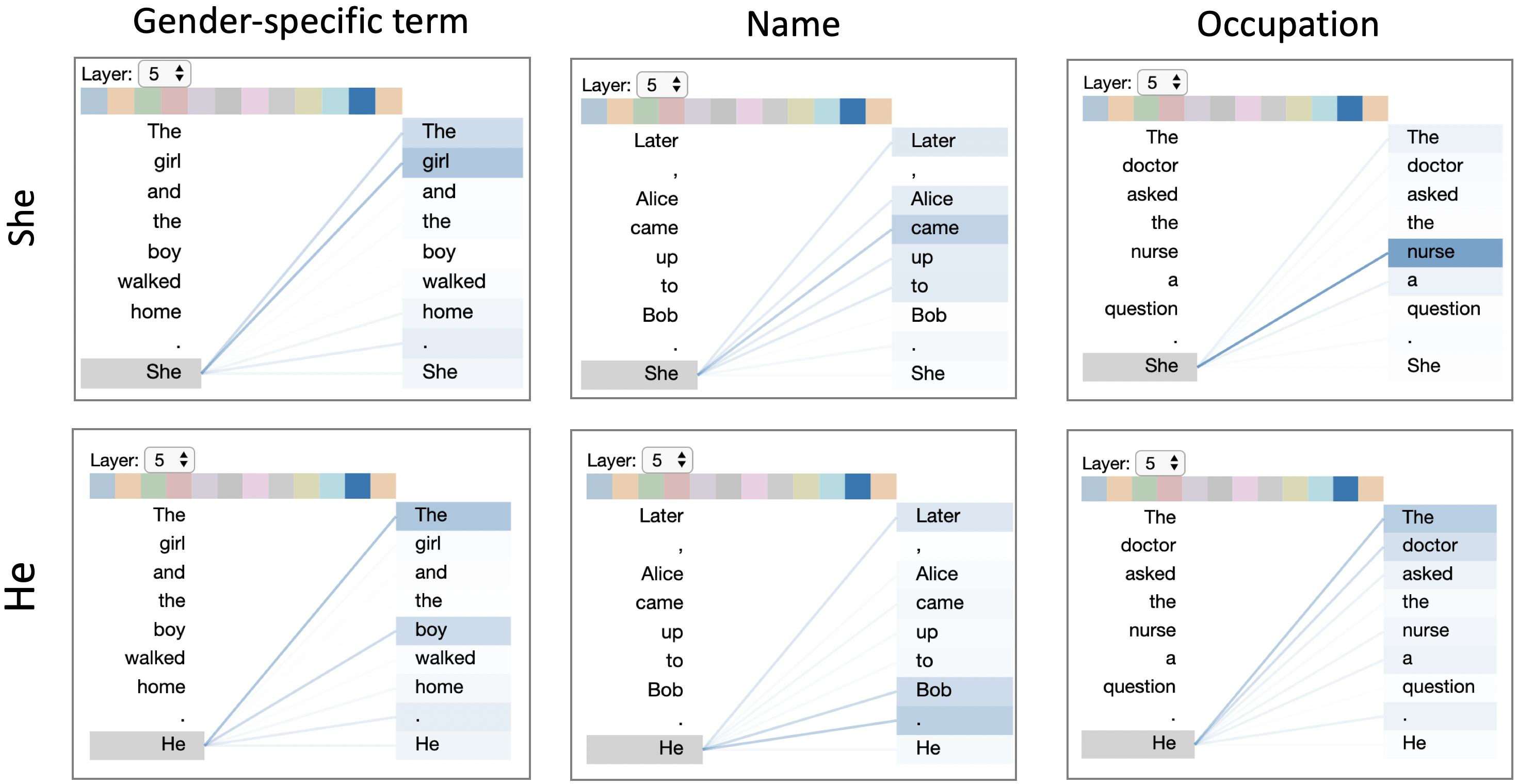}
\caption{Attention pattern in GPT-2 related to coreference resolution suggests the model may encode gender bias.} 
\label{fig:example_bias}
\end{figure*}

The \textit{attention-head view} visualizes the attention patterns produced by one or more attention heads in a given transformer layer, as shown in Figure \ref{fig:head_view_1_combined} (GPT-2\footnote{\textit{GPT-2 small} pretrained model.}) and Figure \ref{fig:bert_heads} (BERT\footnote{$BERT_{BASE}$ pretrained model.}). In this view, self-attention is represented as lines connecting the tokens that are attending (left) with the tokens being attended to (right). Colors identify the corresponding attention head(s), while line weight reflects the attention score.

At the top of the screen, the user can select the layer and one or more attention heads (represented by the colored patches). Users may also filter attention by token, as shown in Figure \ref{fig:head_view_1_combined} (right); in this case the target tokens are also highlighted and shaded based on attention strength. For BERT, which uses an explicit sentence-pair model, users may specify a sentence-level attention filter; for example, in Figure \ref{fig:bert_heads} (right), the user has selected the \textit{Sentence A} $\rightarrow$ \textit{Sentence B} filter, which only shows attention from tokens in Sentence \textit{A} to tokens in Sentence \textit{B}.


Since the attention heads do not share parameters, each head can produce a distinct attention pattern. In the attention head shown in Figure \ref{fig:head_view_1_combined} (left), for example, each word attends exclusively to the previous word in the sequence. The head in Figure \ref{fig:head_view_1_combined} (center), in contrast, generates attention that is dispersed fairly evenly across previous words in the sentence (excluding the first word). Figure \ref{fig:bert_heads} shows attention heads for BERT that capture sentence-pair patterns, including a within-sentence attention pattern (left) and a between-sentence pattern (center).

Besides these coarse positional patterns, attention heads may capture specific lexical patterns such as those shown in Figure \ref{fig:example_combined}. Other attention heads detected named entities (people, places, companies), beginning/ending punctuation (quotes, brackets, parentheses), subject-verb pairs, and a variety of other linguistic properties. 

The attention-head view closely follows the original implementation from \citep{JonesViz}. The key difference is that the original tool was developed for encoder-decoder models, while the present tool has been adapted to the encoder-only BERT and decoder-only GPT-2 models.

\vspace{5px}
\noindent {\bf Use Case: Detecting Model Bias}

\vspace{1px}
\noindent {One use case for the attention-head view is detecting bias in the model. Consider the following two cases of conditional text generation using GPT-2 (generated text underlined), where the two input prompts differ only in the gender of the pronoun that begins the second sentence\footnote{Generated from GPT-2 small model, using greedy top-1 decoding algorithm}:}

\begin{itemize}
\item \textit{The doctor asked the nurse a question. She \textbf{\underline{said, ``I'm not sure what you're talking about.''}}}
\item \textit{The doctor asked the nurse a question. He \textbf{\underline{asked her if she ever had a heart attack.}}}
\end{itemize}

In the first example, the model generates a continuation that implies \textit{She} refers to \textit{nurse}.  In the second example, the model generates text that implies \textit{He} refers to \textit{doctor}. This suggests that the model's coreference mechanism may encode gender bias \cite{gender-bias-in-coreference, gender-bias-nlp}. To better understand the source of this bias, we can visualize the attention head that produces patterns resembling coreference resolution, shown in Figure \ref{fig:example_bias}. The two examples from above are shown in Figure \ref{fig:example_bias} (right), which reveals that the token \textit{She} strongly attends to \textit{nurse}, while the token \textit{He} attends more to \textit{doctor}. This result suggests that the model is heavily influenced by its perception of gender associated with words.

By identifying a potential source of model bias, the tool can help to guide efforts to provide solutions to the issue. For example, if one were able to identify the neurons that encoded gender in this attention head, one could potentially manipulate those neurons to control for the bias \citep{Bau2019}.

\subsection{Model View}

\begin{figure}[h]
    \includegraphics[width=\linewidth]{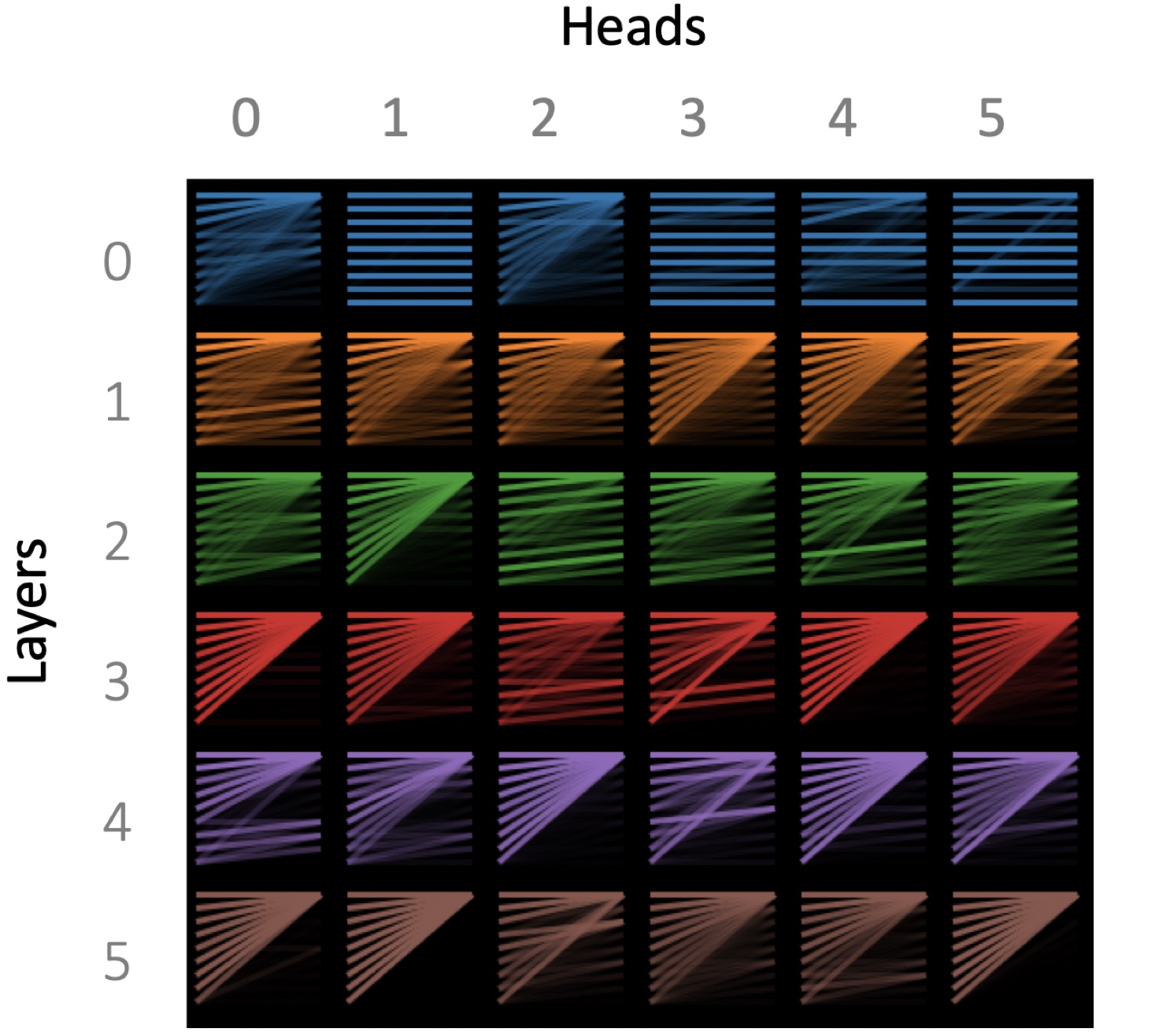}
    \caption{Model view of GPT-2, for input text \textit{The quick, brown fox jumps over the lazy} (excludes layers 6-11 and heads 6-11). }
    \label{fig:model_view}
    
\end{figure}

The \textit{model view} (Figure \ref{fig:model_view}) provides a birds-eye view of attention across all of the model's layers and heads for a particular input. Attention heads are presented in tabular form, with rows representing layers and columns representing heads.  Each layer/head is visualized in a thumbnail form that conveys the coarse shape of the attention pattern, following the \textit{small multiples} design pattern \citep{Tufte1990}. Users may also click on any head to enlarge it and see the tokens.

\begin{figure*}[!t]
    \includegraphics[width=1\linewidth]{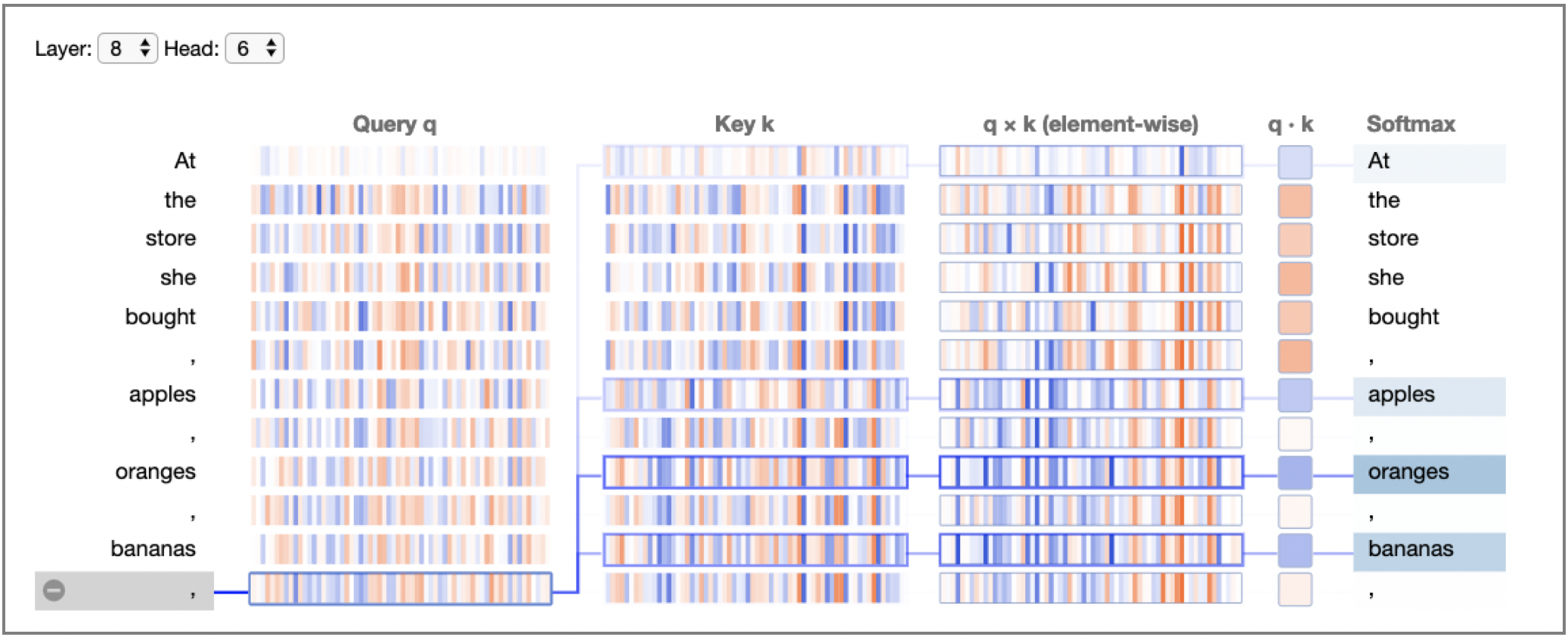}
    \caption{Neuron view of GPT-2 for layer 8 / head 6. Positive and negative values are colored blue and orange respectively, with color saturation reflecting magnitude. This is the same attention head depicted in Figure \ref{fig:example_combined} (left).}
    \label{fig:neuron_view}
\end{figure*}

\begin{figure*}[!b]
    \includegraphics[width=1\linewidth]{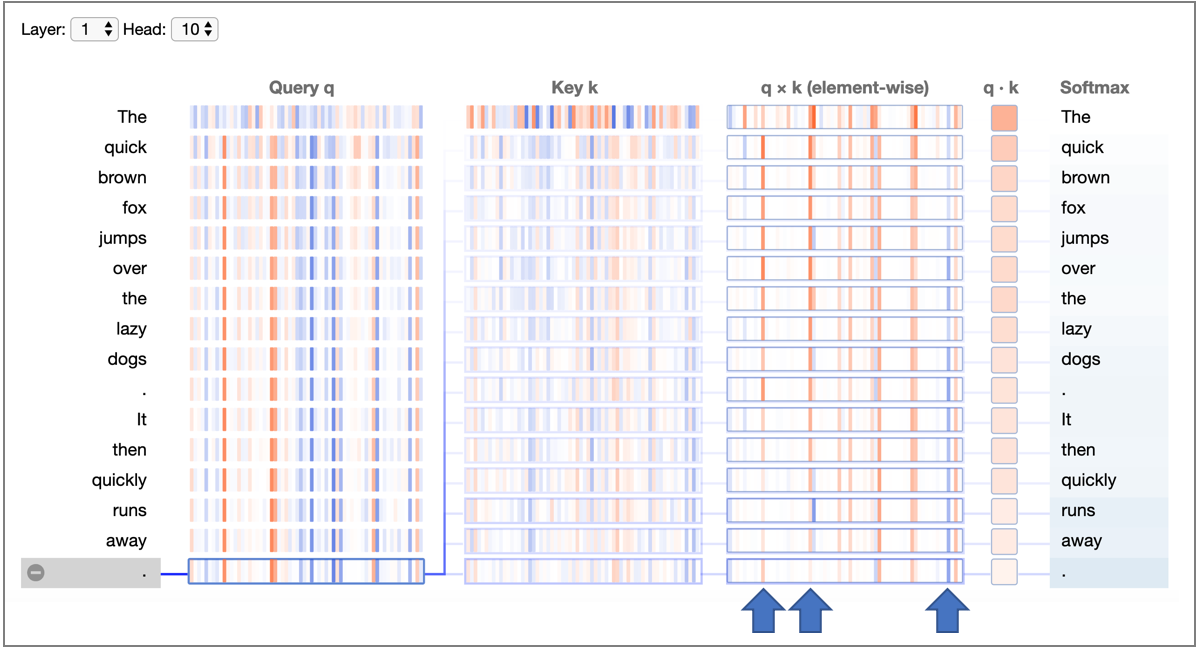}
    \caption{Neuron view of GPT-2 for layer 1 / head 10 (same one depicted in Figure \ref{fig:head_view_1_combined}, center) with last token selected.  Blue arrows mark positions in the element-wise products where values decrease  with increasing distance from the source token (becoming darker orange or lighter blue).}
    \label{fig:neuron_view_2}
\end{figure*}

The model view enables users to browse the attention heads across all layers in the model and see how attention patterns evolve throughout the model. For example, one can see that many attention heads in the initial layers tend to be position-based, e.g. focusing on the same token (layer 0, head 1) or focusing on the previous token (layer 2, head 2).

\vspace{5px}
\noindent {\bf Use Case: Identifying Recurring Patterns}

\vspace{1px}
\noindent {The model view in Figure \ref{fig:model_view} shows that many of the attention heads follow the same pattern: they focus all of the attention on the first token in the sequence. This appears to be a type of null pattern that is produced when the linguistic property captured by the attention head doesn't appear in the input text. One possible conclusion from this result is that the model may benefit from a dedicated null position to receive this type of attention. While it's not clear that this change would improve model performance, it would make the model more interpretable by disentangling the null attention from attention related to the first token.}


\subsection{Neuron View}
The \textit{neuron view} (Figure \ref{fig:neuron_view}) visualizes the individual neurons in the query and key vectors and shows how they are used to compute attention. Given a token selected by the user (left), this view traces the computation of attention from that token to the other tokens in the sequence (right). The computation is visualized from left to right with the following columns:
\begin{itemize}
    \item \textbf{Query q}: The 64-element query vector of the token paying attention. Only the query vector of the selected token is used in the computations. 
    \item \textbf{Key k}: The 64-element key vector of each token receiving attention. 
    \item \textbf{q $\times$ k (element-wise)}: The element-wise product of the selected token's query vector and each key vector. 
    \item \textbf{q $\cdot$ k}: The dot product of the selected token's query vector and each key vector.
    \item \textbf{Softmax}: The softmax of the scaled dot-product from previous column. This equals the attention received by the corresponding token.
\end{itemize}

Positive and negative values are colored blue and orange, respectively, with color saturation based on the magnitude of the value. As with the attention-head view, the connecting lines are weighted based on attention between the words. The element-wise product of the vectors is included to show how individual neurons contribute to the dot product and hence attention. 

\vspace{5px}
\noindent {\bf Use Case: Linking Neurons to Model Behavior}

\vspace{1px}
\noindent {To see how the neuron view might provide actionable insights, consider the attention head in Figure \ref{fig:neuron_view_2}. For this head, the attention (rightmost column) appears to decay with increasing distance from the source token\footnote{with the exception of the first token, which acts as a null token, as discussed earlier.}. This pattern resembles a context window, but instead of having a fixed cutoff, the attention decays continuously with distance.}

The neuron view provides two key insights about this attention head. First, the attention scores appear to be largely independent of the content of the input text, based on the fact that all the query vectors have very similar values (except for the first token). The second observation is that a small number of neuron positions (highlighted with blue arrows) appear to be mostly responsible for this distance-decaying attention pattern. At these neuron positions, the element-wise product q$\times$k decreases as the distance from the source token increases (either becoming darker orange or lighter blue).



When specific neurons are linked to a tangible outcome---in this the case decay rate of attention---it presents an opportunity for human intervention in the model. By altering the values of the relevant neurons \citep{Bau2019}, one could control the rate at which attention decays for this attention head. This capability might be useful when processing or generating texts of varying complexity; for example, one might prefer a slower decay rate (longer context window) for a scientific text and a faster decay rate (shorter context window) for content intended for children.


    
    

\section{Conclusion}
In this paper, we presented a tool for visualizing attention in Transformer-based language representation models. We demonstrated the tool on the OpenAI GPT-2 and BERT models and presented three use cases for analyzing GPT-2. For future work, we would like to evaluate empirically how attention impacts model predictions across a range of tasks \citep{attention_not_explanation}. Further, we would like to integrate the three views into a unified interface, and visualize the value vectors in addition to the queries and keys. Finally, we would like to enable users to manipulate the model, either by modifying attention \citep{Lee2017, visual_interrogation, seq2seqvisv1} or editing individual neurons \citep{Bau2019}. 

\bibliography{main}
\bibliographystyle{acl_natbib}

\end{document}